\documentclass[aps,prc,twocolumn,showpacs,amsmath,amssymb]{revtex4}

\usepackage{graphicx}

\begin{document}


\title{Resonant State in $^{\mathbf 4}_{\mathbf\Lambda}$He} 

\author{D.~E.~Kahana}
\author{S.~H.~Kahana}
\author{D.~J.~Millener}
 
\affiliation{Physics Department, Brookhaven National Laboratory,
   Upton, NY 11973, USA}

\date{\today}  
 
\begin{abstract}

In a recent experiment $E906$  at the BNL-AGS, a search for light
$S=-2$  hypernuclei,  strong evidence  for  the  nuclide $^{\  \:
4}_{\Lambda\Lambda}$H  was  found.   One  of  the  most  striking
components  of   this  data  was  the  appearance   of  a  narrow
low-momentum $\pi^{-}$  line at  $k_{\pi} = 104-105$  MeV/c. This
was ascribed to the decay of $^{\ \: 4}_{\Lambda\Lambda}$H into a
resonant state in $^4_{\Lambda}$He. The existence of such a state
is shown to be plausible and its characteristics delineated.

\end{abstract}


\maketitle

Several    candidates   \cite{Danysz,Prowse,Aoki,takahashi}   for
doubly-strange  hypernuclei  have  been  identified  in  emulsion
experiments.  By the nature  of the latter approach, usually only
one example of a given  species is uncovered in each search.  The
BNL-AGS experiment E906 \cite{e906}, a counter experiment looking
for correlated two $\pi^-$ decays, was able to accumulate several
tens   of  plausible   samples  of   the  hypernucleus   $^{\  \:
4}_{\Lambda\Lambda}$H.  In this work we concentrate on one of the
prominent    decay    modes    of    this    nuclide,    ascribed
\cite{e906,abstract} to  initial decay  into a resonant  state in
$^4_{\Lambda}$He. This  is in analogy to the  well-known decay of
$^5_\Lambda$He which  gives rise to  a rather narrow peak  in the
$\pi^-$  spectrum~\cite{Li5}. We  simply extend  the  analysis of
E906 in a most natural fashion.

One   expects   the   major   sequential  decays   of   $^{\   \:
4}_{\Lambda\Lambda}$H to be:

\begin{eqnarray}
   ^{\  \:  4}   _{\Lambda\Lambda}\text{H}  &  \rightarrow  &  \,
_{\Lambda}^4\text{He}   +   \pi^-_H   \quad\quad  (\sim   112-118~
\text{MeV/c})  \\   ^4_{\Lambda}\text{He}  &  \rightarrow   &  \
^{3}\text{H} + p + \pi^-_L\quad\ (\sim 85-95 ~\text{MeV/c})
\end{eqnarray}
and  in particular,  a decay  into  a possible  excited state  of
$_{\Lambda}^4$He,
\begin{eqnarray}
   _{\Lambda\Lambda}^{\   \:  4}\text{H}   &  \rightarrow   &  \,
_\Lambda^4\text{He}^* + \pi^-_L\quad (\sim 104-105 ~\text{MeV/c})
\label{eq:pi1}\\
_\Lambda^4\text{He}^* &  \rightarrow & \, _\Lambda^3\text{H}
+  p \label{eq:proton}\\ 
 _\Lambda^3\text{H} &  \rightarrow &  \,^{3}\text{He} +
\pi^-_H \quad\quad (114.3 ~\text{MeV/c}), \label{eq:pi2}
\end{eqnarray}
where $\pi^-_H$ and $\pi^-_L$ refer  to the high and low momentum
members  of  a correlated  pair  seen  in  the experiment. 

The resonance  depicted in Eqs.~(3-4), as  indicated below, would
certainly  be  suppressed  in  the  standard  single  hypenucleus
search,        for       example       in        the       simple
$^4\text{He}(K^-,\pi^-){^4_\Lambda}$He     reaction~\cite{kalpha}.
Nevertheless,  such a  state plays  a dominant  role in  the E906
study, seemingly the  only explanation~\cite{e906} for the strong
$\pi^-$ decay line seen at  $104-105$ MeV/c.  Such a state can be
shown  to  be theoretically  plausible,  arising  naturally in  a
model~\cite{abstract} consistent  with the known  ground state of
$_{\Lambda}^4$He.  The  approach followed  here is to  identify a
candidate, not surprisingly a (proton + $^3_{\Lambda}$H) relative
p  state,  and  establish  its  suitability.   A  state  of  this
architecture   would    lead   directly   via    the   decay   of
$^3_{\Lambda}$H, to  the higher momentum $\pi^-$  decay line near
114 MeV/c  correlated with  the considerably narrower  feature at
$104-105$ MeV/c in the E906 data~\cite{e906}.

As was demonstrated in  Ref.~\cite{kumagai02}, this state has the
characteristics for  being a  conspicuous daughter for  the decay
mode  cited above  in  Eq.~(3).   As we  indicate  later, we  can
construct  simple non-spurious $1\hbar\omega$  shell-model states
which should be  populated strongly in the $\pi^-$  decay of $^{\
\: 4}_{\Lambda\Lambda}$H, and  should themselves decay via proton
rather  than $\Lambda$  emission.  Ideally  one might  attempt to
treat   the    specific   4-body   nature   of    the   $^{\   \:
4}_{\Lambda\Lambda}$H             and            $^4_{\Lambda}$He
systems~\cite{kumagai02,Gal,Akaishi}.   However,  practically one
is then limited to  a restricted selection of possible two-baryon
interactions.  Even in the $S=-1$ baryon-baryon sector, the known
forces, constrained  by available data, are by  no means uniquely
defined, and  hence perhaps as  yet unreliable. To  establish the
presence  of  a  weakly  bound,  or  in  particular  a  low-lying
resonant,  state is  then a  daunting theoretical  task.  Indeed,
existing  many body  calculations~\cite{3LHcalc}  for the  weakly
bound $^3_{\Lambda}$H indicate that $\Lambda$N-$\Sigma$N coupling
must   be  included,   as   it  must   for   all  the   $s$-shell
hypernuclei~\cite{sshell}.  This  adds to the  complexity of such
calculations.   Ultimately,  it  is  precisely  the  experimental
evidence considered here and  elsewhere that should constrain and
illuminate hypernuclear few-body theory.

At present, as already  indicated, reasonable evidence exists for
such  a state  from experiment~\cite{e906}.   The dynamics  of an
essentially  (p +  $^3_{\Lambda}$H) p-state  can perhaps  be well
approximated  by a  one body  + cluster  potential  problem.  The
effective  interaction  between a  proton  and  a three  particle
hyperon core can be viewed  as the average potential arising {\it
after}  a combined  many-body,  configuration mixing,  treatment.
One is  here essentially only  interested in the basic  s-wave to
p-wave   separation    obtaining,   very   much    an   effective
single-particle property.  That the anticipated resonance emerges
naturally from such modeling is suggestive.

The characteristics of such a  potential are fairly clear, one is
dealing  with  rather spatially  extended  systems  for both  the
weakly  bound   $^{\  \:  4}   _{\Lambda\Lambda}$H  and  resonant
$^4_\Lambda$He$^*$.   Moreover,  the  potential can  be  strongly
constrained, by  the requirement that its  depth also accommodate
the  ground  state of  $^4_{\Lambda}$He.   The spatial  extension
suggests  one  might use  a  potential  with  a relatively  large
radius,  or even  a surface  potential.  The  resulting spatially
large state also  favors the decay into the  resonance in Eq.~(3)
from the expected weakly bound $^{\ \: 4} _{\Lambda\Lambda}$H.
 
To  help pick  out  possible candidates  for the  negative-parity
resonance in $^4_\Lambda$He, and to illustrate important features
of the $\pi^-$  weak decay of $^{\ \:  4} _{\Lambda\Lambda}$H, we
enumerate  in   Eqs.~(\ref{basis1})-(\ref{basis3})  the  possible
1$\hbar\omega$,  $T=1/2$,   states  in  an   harmonic  oscillator
shell-model  basis  with  $\hbar\omega_N =  \hbar\omega_\Lambda$.
Although  harmonic  oscillator  radial  wave  functions  are  not
appropriate for the loosely  bound or resonant states, this basis
nevertheless  has the  virtue that  the spurious  center  of mass
states  can  be eliminated,  as  in  Eq.~(\ref{basis1}), and  the
explicit intrinsic-spin  structure of  the states makes  it clear
which states  can be  fed strongly in  the $\pi^-$ weak  decay of
$^{\ \: 4} _{\Lambda\Lambda}$H.
\begin{eqnarray}
 \sqrt{\frac{\mu}{3+\mu}}  &  (s^2p)[3]1/2\times  s_\Lambda &  -\
\sqrt{\frac{3}{3+\mu}}\   s^3\times  p_\Lambda\label{basis1}\\  &
(s^2p)[21]1/2\times     s_\Lambda     &    \label{basis2}\\     &
(s^2p)[21]3/2\times s_\Lambda & \label{basis3}
\end{eqnarray}
Here $\mu  = m_\Lambda/m_N$ and [$f$]$S_{3N}$  labels the spatial
symmetry and  intrinsic spin of  the three nucleons. For  each of
the  three  classes of  states  $L=1$  and  two values  of  total
intrinsic spin $S$ are possible.   In the same model, the $^{\ \:
4}  _{\Lambda\Lambda}$H  initial  state is  simply  $s^2(1)\times
s^2_\Lambda(0)$ with $S  = 1$. Because the dominant  piece of the
weak-decay  operator does  not  involve intrinsic  spin, we  need
consider  only $^4_\Lambda$He  states with  $L =  1$,  $S=1$, and
$J=0,1,2$.

\begin{figure}
\includegraphics[width=3.1in]{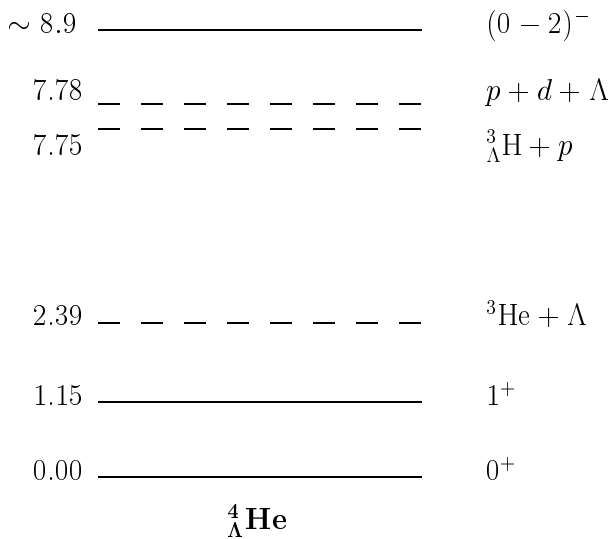}
\caption{\label{fig:4lhe}  Spectrum   and  decay  thresholds  for
$^4_\Lambda$He.  The postulated  negative-parity  resonant states
are indicated above the proton threshold.}
\end{figure}

Because  the proton threshold  in $^4_\Lambda$He  at 7.75  MeV is
much  higher  than  the  $\Lambda$  threshold at  2.39  MeV  (see
Fig.~\ref{fig:4lhe})  any  state with  even  a small  $p_\Lambda$
component  will decay  to $^3$He$+\Lambda$.   This rules  out the
states of  Eq.~(\ref{basis1}), although such states  could be fed
via  the $s_\Lambda$  component in  the weak  decay leading  to a
$^3$He$+\Lambda$ final  state. The states  in Eqs.~(\ref{basis2})
and  (\ref{basis3}),  which   should  be  at  similar  excitation
energies, qualify as candidates.  We do not expect the relatively
weak $\Lambda$N interaction  to mix these configurations strongly
with those of Eq.~(\ref{basis1}).  We note that in light nuclei a
number of narrow states exist at high excitation energies because
of the symmetry  structure of the wave functions,  e.g. a $3/2^+$
state of $t+d$ structure at 16.75 MeV, in $^5$He and an analogous
state in $^5$Li  despite a large decay energy  into the $\alpha +
N$  channel.  If  the intrinsic  spins in  the  configurations in
Eqs.~(\ref{basis2}) and (\ref{basis3})  are recoupled to the form
$(s^2s_\Lambda)S_{3H}\times  p$, $S_{3H} =  1/2$ is  required for
the $^3_\Lambda\text{H}+ p$ channel;  $S_{3H} = 3/2$ leads to the
$d  +\Lambda +  p$ final  state.  In  fact, the  configuration in
Eq.~(\ref{basis3})  dominates the  $S_{3H} =  1/2$  strength. The
preceding discussion  relating to the possible  decay channels of
single-$\Lambda$  hypernuclear  configurations  produced  in  the
$\pi^-$  weak   decay  of  $^{\  \:   4}  _{\Lambda\Lambda}$H  is
summarized in rows $2-5$ of Table~\ref{tab:decay}.

The weak-decay of $^{\  \: 4}_{\Lambda\Lambda}$H has been treated
by Kumagai-Fuse and  Okabe~\cite{kumagai02}. The widths that they
calculate   for   decays    to   $^4_\Lambda$He$(1^+)$   and   to
$^3_\Lambda$H$+p$        are       0.68$\Gamma_\Lambda$       and
0.80$\Gamma_\Lambda$,  respectively.  We reproduce  the essential
features of  these results using  just the dominant piece  of the
weak-decay operator  which does not involve the  nucleon spin. In
the  plane-wave limit for  the outgoing  $\pi^-$ the  operator is
simply $s_{\pi^-}j_\lambda(k_\pi  r)Y_\lambda$, where $s_{\pi^-}$
is   a   strength   known    from   the   decay   of   the   free
$\Lambda$~\cite{kumagai02}. The squares of the matrix elements of
this operator  taking into  account the sum  over $J_f$  with the
usual statistical weighting  of $(2J_f+1)/(2J_i+1)$ (Eqs.~(9) and
(10) of  \cite{kumagai02})   are  given  in  the   first  row  of
Table~\ref{tab:decay}.  Putting in the kinematic factors (and the
small  contribution from  the spin-dependent  weak-decay operator
with  a strength given  by $p_{\pi^-}$~\cite{kumagai02})  gives a
width                                    $0.806\left<s_N|j_0(k_\pi
r)|s_\Lambda\right>^2\Gamma_\Lambda$ to the $^4_\Lambda$He$(1^+)$
state, where the  radial matrix element is not  very much smaller
than  unity because  $k_\pi\sim 0.53$  fm$^{-1}$ is  quite small.
The   radial   matrix  element   for   the   decays  leading   to
$^3_\Lambda$H$+p$ is  smaller and a rough  estimate suggests that
it compensates  for the 11/3  factor for production noted  in the
caption  of Table~\ref{tab:decay}  to give  a rate  comparable to
that  for the  $^4_\Lambda$He$(1^+)$  state.  The  point is  that
there is quite good agreement with Ref.~\cite{kumagai02} and that
$^3_\Lambda$H   should   be   produced  sufficiently   that   the
scenario~\cite{e906}   put    forward   in   Eqs.~(\ref{eq:pi1}),
(\ref{eq:proton}), and (\ref{eq:pi2}) is reasonable.

\begin{table}
\caption{\label{tab:decay} Matrix elements  for the production of
single-$\Lambda$ hypernuclear configurations  in the $\pi^-$ weak
decay of  $^{\ \: 4}  _{\Lambda\Lambda}$H are given in  the first
row     in    units     of    $s_{\pi^-}^2\left<l_N|j_{l_N}(k_\pi
r)|s_\Lambda\right>^2/4\pi$   where  $\mu/(3+\mu)=   0.284$.  The
remaining  rows  specify  the  breakup  of  the  single-$\Lambda$
hypernuclear  configurations under  the assumptions  described in
the text. Combining production  and decay in this simple approach
shows that  $^3_\Lambda$H$+p$ is favored over  $d+p+\Lambda$ by a
factor of 11/3 to 4/3.}
\begin{ruledtabular}
\begin{tabular}{ccccc}
 Final state & $^4_\Lambda$He$(1^+)$ & Eq.~(\ref{basis1}) &
 Eq.~(\ref{basis2})  & Eq.~(\ref{basis3}) \\
\hline
 Production & 1 & $\mu/(3+\mu)$ & 1 & 4 \\
$^4_\Lambda$He$(1^+)$ & 1 & & & \\
$^3$He$+\Lambda$ & & 1 & & \\
$^3_\Lambda$H$+p$ & & & 1/9 & 8/9 \\
$d+p+\Lambda$ & & & 8/9 & 1/9 \\
\end{tabular}
\end{ruledtabular}
\end{table}

The next order  of business is to construct  an average potential
describing a  proton in  both the ground  and resonant  states of
$^4_{\Lambda}$He.   We  reemphasize  that  we are  using  such  a
potential to represent the average interaction between the proton
and an effective $^3_{\Lambda}$H  cluster.  We employ a resonance
code due  to T.~Vertse, K.~F.~Pal  and Z.~Balogh~\cite{resonance}
to perform  the necessary  calculations, together with  a surface
Saxon-Woods  potential. The  addition of  a  spin-orbit potential
doesn't  change   the  basic  results  and,  in   any  case,  the
configurations   in    Eqs.~(\ref{basis2})   and   (\ref{basis3})
correspond to mixtures of  $p_{1/2}$ and $p_{3/2}$ proton orbits.
Thus, we take,
\begin{equation}
V^{Surf}(r)= -V_0\left[\frac{4e^{(r-R)/a}}{(1+e^{(r-R)/a})
^2}\right],
\end{equation}
with   $R=r_0  A^{1/3}$   the  potential   radius  and   $a$  the
diffusivity.  A Coulomb  potential, important to constraining the
width of a resonant proton  line, is included.  This potential is
taken  as  that of  a  uniform  charge  distribution with  radius
parameter $r_{0c}=r_0$.

The ground state proton has  of course $l=0$ and a known binding.
Only the  strength of the  surface potential component  is varied
and  reproduces the  correct 7.75  MeV separation  energy  with a
depth  $V_0=28.09$  MeV,  radius  parameter  $r_0=1.40$  fm,  and
diffusivity  $a=0.5$  fm. It  should  be  noted  that the  actual
nuclear radius is  $R=2.07$ fm since the A  used in the resonance
code~\cite{resonance}  is  the   hypernuclear  core  mass  $3.21$
amu. In  this completely  specified well, the  $p$-wave resonance
appears at $\epsilon=1.18$ MeV with  a width of $1.00$ MeV.  Both
these  numbers are  consistent with  the observed  narrow $\pi^-$
line at $104-106$ MeV/c in E906, recalling that the resolution in
this experiment  was $\sim 2.5$  MeV.  For the  choice $r_0=1.45$
fm,   the  corresponding  well   and  resonance   parameters  are
$V_0=27.68$, $\epsilon=0.97$ MeV, and width $0.70$ MeV.

Another possible variation is to fix the proton separation energy
at $7.18$  MeV, from averaging the  $^4_{\Lambda}$He ($0^+, 1^+$)
ground state  doublet positions. Resetting  the radius parameters
to $r_0=1.45$ fm, the p-wave resonance appears at $1.19$ MeV with
a width of  $1.04$ MeV in a slightly  shallower surface well with
$V_0=26.69$ MeV.  One is,  of course, comparing only to centroids
of the experimental features.

Most  convincing in  this analysis  is  the ease  with which  the
$^4_{\Lambda}$He  resonance is extracted,  once the  ground state
binding of the proton is assigned. The final resonance parameters
obtained cannot be taken too  literally, but the energy and width
have  little   variability  with  reasonable   changes  in  other
potential parameters.

With the resonance energy calculated the $\Lambda\Lambda$ pairing
energy  $\Delta\text{B}_{\Lambda\Lambda}$ can  be  estimated from
the  position of  the narrow  $\pi^-$ peak  ascribed to  the weak
decay     from    $^{\     \:    4}     _{\Lambda\Lambda}$H    to
$^4_{\Lambda}\text{He}^*$. The $\pi^-$ momentum $k^*$ is given by
\begin{equation}
\epsilon_{\pi}+k^{*2}/2M(^4_{\Lambda}\text{He}^*)    =   M(   ^{\
\:4}_{\Lambda\Lambda}\text{H})-M(^4_{\Lambda}\text{He}^*).
\label{eq:pionkf}
\end{equation}
To  a good  approximation in  the region  of  momentum considered
here,
\begin{equation}
 k_{\pi}^* = (107.466-1.6391\Delta) ~\text{MeV/c},  
\label{eq:pionka}
\end{equation}
where     $\Delta=\text{B}^*    +    \epsilon_R$     and    B$^*=
2{\bar{\text{B}}}_{\Lambda}(^3_{\Lambda}\text{H})+
\Delta\text{B}_{\Lambda\Lambda}$  is the  full binding  energy of
the  $\Lambda$  pair  in  $^{\ \:  4}  _{\Lambda\Lambda}\text{H}$
excluding  rearrangement effects  and coupling  to  other hyperon
channels.  These relations exhibit the dependence of the measured
meson   momentum   on   the   combined   resonance   energy   and
$\Lambda\Lambda$   interaction   energy.   In   evaluating   the
contribution of  the $\Lambda$N interaction, a  recoupling of the
initial     $s^2\times     s^2_\Lambda$     wavefunction     into
$(s^2s_\Lambda)S_{3H}$ coupled  to a spectator  $s_\Lambda$ leads
to  a statistical weighting  of B$_\Lambda$  values for  the real
$1/2^+$   hypertriton  and  the   unbound  and   unknown  $3/2^+$
hypertriton  denoted  by  ${\bar{\text{B}}}_\Lambda$.  Using  the
known  value for  2B$_\Lambda(^3_{\Lambda}\text{H})$ in  place of
2${\bar{\text{B}}}_\Lambda$   as   we   do   below   means   that
$\Delta\text{B}_{\Lambda\Lambda}$ will be underestimated (but not
by a large amount).  In our model, for example, taking $r_0=1.40$
fm,       hence       $\epsilon_R=       1.18$      MeV       and
$\Delta\text{B}_{\Lambda\Lambda}=0.34$ MeV,  puts the centroid of
the decay  momentum at $k^*=104.5$ MeV/c,  nominally the position
of the measured centroid  for this decay momentum. Alternatively,
$\epsilon_R=0.97$  MeV and  the  same decay  momentum results  in
$\Delta\text{B}_{\Lambda\Lambda}$= 0.55 MeV.

The indefiniteness  in the measured  $\pi_L$ momentum \cite{e906}
subjects the  pairing energy  to some uncertainty.   For example,
with $\epsilon_R=1.18$,  lowering $k^*$  by $300$ keV  results in
$\Delta\text{B}_    {\Lambda\Lambda}=0.55$     MeV,    while    a
corresponding increase in $k^*$  yields $0.17$ MeV.  Whatever the
choice, $\Delta\text{B}_{\Lambda\Lambda}$  here remains small and
appreciably less than the $\sim  1$ MeV determined for $^{\ \: 6}
_{\Lambda\Lambda}$He \cite{takahashi}.

One  very  interesting  feature  of these  estimated  values  for
$\Delta\text{B}_{\Lambda\Lambda}$,   say  $0.34$   MeV,   is  the
momentum for the truly two body decay
\begin{equation}
 ^{\    \:    4}    _{\Lambda\Lambda}\text{H}   \rightarrow    \,
_{\Lambda}^4\text{He}(1^+) + \pi^-_H.
\end{equation}
A  similar approximation  to that  in Eq.~(\ref{eq:pionka})  
produces,  with no $\epsilon_R$ involvement,
\begin{equation}
 k_{\pi}(1^+) = (117.454-1.5531B^*) ~\text{MeV/c},
\end{equation}
and  $k_{\pi}  =  116.5$  MeV/c.   Such a  line,  although  still
consistent with the observed  broad peak near $115$ MeV/c, should
be  relatively  easy  to  separate,  in  an  improved  resolution
experiment, from the known $114.3$ MeV/c for the decay
\begin{equation}
 ^3_{\Lambda}\text{H} \rightarrow \, ^3\text{He} + \pi^-_H\ .
\end{equation}
Finding this  peak in the  $\pi^-_H$ spectrum would  provide more
transparent   evidence  for   the   existence  of   $^{\  \:   4}
_{\Lambda\Lambda}\text{H}$.

In  conclusion,  we  see  that  the  existence  of  states  above
threshold    in    $_{\Lambda}^3$H    +   proton    system    can
straightforwardly  describe  the  narrow, low  momentum,  $\pi^-$
feature  at  $104-105$  MeV/c  observed  in  the  BNL  experiment
E906. An  argument in favor of  such resonances is  that a common
average particle-cluster  potential can  be used to  describe the
gross   structure   of   both   the   ground-state   doublet   of
$^4_\Lambda$He and the  negative-parity resonances. To pursue the
problem  further, one  must surely  involve the  full interaction
between  the  ``resonant''  proton  and  its  hypernuclear  core,
including the NN, both the  odd and even state $\Lambda$N, forces
and    very    likely    $\Lambda$N-$\Sigma$N    coupling.     In
Ref.~\cite{kumagai02}, a  p-wave resonance could  not be produced
by varying  only the poorly known, and  probably relatively weak,
odd-state  $\Lambda$N  strength.   These very  light  hypernuclei
certainly warrant more theoretical study.

The inferred  energy of the  resonance and energy  range possible
for $\Delta\text{B}_{\Lambda\Lambda}$, the latter likely somewhat
less  than 0.5  MeV,  are not  unreasonable.   Takahashi {\it  et
al.}~\cite{takahashi} found $\Delta\text{B}_{\Lambda\Lambda}= 1.0
\pm 0.38$ MeV for  $_{\Lambda\Lambda}^{\ \: 6}$He.  One certainly
expects a smaller value for  the more extended mass 4 system, but
the  suggested   values  and   their  likely  errors   allow  for
consistency.  Only future  experiments with better statistics and
better resolution  can settle  this issue.  We  reemphasize that,
although  the   analysis  followed  here  is   the  most  natural
interpretation   of   the   existing   data   \cite{e906},   this
interpretation  would  be   much  illuminated  by  such  followup
experiments.

For  the meantime,  it  is  also of  high  importance to  perform
many-body  calculations~\cite{kumagai02,Gal,Akaishi}  with forces
constrained by as much data  as is available, including of course
that  from  E906.   The  latter  experiment has  pointed  to  the
existence of two interesting nuclides, both mass 4 objects, one a
very  light $S=-2$  hypernucleus, the  other an  unusual,  if not
completely unexpected, resonance in an $S=-1$ daughter nucleus.
   
The         initial         discoveries         of         $S=-2$
nuclei~\cite{Danysz,Prowse,Aoki}  had one measurement  in common,
the $\sim  4.5$ MeV value  for $\Delta\text{B}_{\Lambda\Lambda}$,
irrespective   of  species.    This  seemed   excessively  large,
considering  the expectation  of a  rather  weak $\Lambda\Lambda$
interaction,  and  hinted  at  possibly interesting  short  range
behaviour,   perhaps   even    bag-like   structure,   in   these
doubly-strange  hypernuclei.  The  value implied  by  the present
analysis   and   by    the   recent   KEK   emulsion   experiment
\cite{takahashi}, $\Delta\text{B}_{\Lambda\Lambda} \le 1.4 $ MeV,
apparently put paid to such speculation. However, the possibility
of unusual short range  structure cannot be completely ruled out,
and would  certainly complicate the calculation  of light systems
containing  a   $\Lambda$  pair.   Incidentally,   a  judiciously
selected   excited  state  inserted   into  the   early  analyses
\cite{Danysz,Aoki,Millener} would also lower the pairing energies
extracted  there, in line  with the  recent measurements  and the
present analysis.

One      can      also       infer,      from      the      small
$\Delta\text{B}_{\Lambda\Lambda}$ suggested  here for $^{\  \: 4}
_{\Lambda\Lambda}\text{H}$  ,  a possible  binding  of less  than
$1.0$  MeV  for  the  highly  elusive  $H$-dibaryon~\cite{Jaffe}.
Production  of this  object  is severely  reduced  by the  strong
likelihood   of  a   repulsive  core   in   the  $\Lambda\Lambda$
interaction \cite{kahana2}.   Evidently, the threshold  for study
of  $S=-2$  nuclei  has   just  been  broached;  many,  and  more
extensive, searches are needed.

This  manuscript  has  been  authored  under  the  US  DOE  grant
NO. DE-AC02-98CH10886.

\end{document}